\documentclass[twocolumn,superscriptaddress,aps,prr,nofootinbib,floatfix]{revtex4-2}
\usepackage{amssymb}
\usepackage{amsmath}
\usepackage{graphicx}
\usepackage{dcolumn}
\usepackage{bm}
\usepackage{dsfont}
\usepackage{CJKutf8} 
\usepackage[breaklinks=true,colorlinks,citecolor=blue,linkcolor=blue,urlcolor=blue]{hyperref}

\usepackage[caption=false]{subfig}
\usepackage{physics}
\usepackage{mleftright}
\usepackage{bbm}
\usepackage{soul}
\usepackage{enumitem}

\usepackage{tikz,xcolor}
\definecolor{lime}{HTML}{A6CE39}
\DeclareRobustCommand{\orcidicon}{%
	\begin{tikzpicture}
	\draw[lime, fill=lime] (0,0) 
	circle [radius=0.16] 
	node[white] {{\fontfamily{qag}\selectfont \tiny ID}};
	\draw[white, fill=white] (-0.0625,0.095) 
	circle [radius=0.007];
	\end{tikzpicture}
	\hspace{-2mm}
}
\foreach \x in {A, ..., Z}{%
	\expandafter\xdef\csname orcid\x\endcsname{\noexpand\href{https://orcid.org/\csname orcidauthor\x\endcsname}{\noexpand\orcidicon}}
}

\begin{document}
\begin{CJK*}{UTF8}{bsmi}
\title{Quantum heat engine based on quantum interferometry: the SU$(1,1)$ Otto cycle}
\date{\today}
\author{Alessandro Ferreri\orcidA{}}
\email{a.ferreri@fz-juelich.de}
\affiliation{Institute for Quantum Computing Analytics (PGI-12), Forschungszentrum J\"ulich, 52425 J\"ulich, Germany}
\affiliation{Theoretical Quantum Physics Laboratory, RIKEN, Wako-shi, Saitama 351-0198, Japan}
\author{Hui Wang (王惠) \orcidE{}}
\affiliation{Theoretical Quantum Physics Laboratory, RIKEN, Wako-shi, Saitama 351-0198, Japan}
\author{Franco Nori \mbox{(野理)} \orcidD{}}
\affiliation{Theoretical Quantum Physics Laboratory, RIKEN, Wako-shi, Saitama 351-0198, Japan}
\affiliation{Center for Quantum Computing, RIKEN, Wako-shi, Saitama 351-0198, Japan}
\affiliation{Physics Department, The University of Michigan, Ann Arbor, Michigan 48109-1040, USA}
\author{Frank K. Wilhelm\orcidC{}}
\affiliation{Institute for Quantum Computing Analytics (PGI-12), Forschungszentrum J\"ulich, 52425 J\"ulich, Germany}
\affiliation{Theoretical Physics, Universit\"at des Saarlandes, 66123 Saarbr\"ucken, Germany}
\author{David Edward Bruschi\orcidB{}}
\email{david.edward.bruschi@posteo.net}
\email{d.e.bruschi@fz-juelich.de}
\affiliation{Institute for Quantum Computing Analytics (PGI-12), Forschungszentrum J\"ulich, 52425 J\"ulich, Germany}
\affiliation{Theoretical Physics, Universit\"at des Saarlandes, 66123 Saarbr\"ucken, Germany}
\begin{abstract}
We present a quantum heat engine based on a quantum Otto cycle, whose working substance reproduces the same outcomes of a SU$(1,1)$ interference process at the end of each adiabatic transformation. This device takes advantage of the extraordinary quantum metrological features of the SU$(1,1)$ interferometer to better discriminate the sources of uncertainty of relevant observables during each adiabatic stroke of the cycle. Applications to circuit QED platforms are also discussed.
\end{abstract}
\maketitle
\end{CJK*}
\section*{Introduction}
The advent of quantum thermodynamics marks the beginning of a modern way of conceiving the laws of thermodynamics \cite{doi:10.1080/00107514.2016.1201896,kurizki2022thermodynamics,e15062100,binder2018thermodynamics,potts2024quantumthermodynamics,deffner2019quantum,RevModPhys.81.1}. Whereas classical thermodynamics relies its predictions on the statistical behavior of large systems characterized by an uncountable number of components (e.g. the molecules of a gas confined in a piston), quantum thermodynamics studies concepts such as heat transport, entropy production, and work extraction at the quantum scale. 
One of the main goals within this framework is to miniaturize heat engines by utilizing standard quantum systems, such as qubits or quantum harmonic oscillators, as working substance \cite{quan_quantum_2007,quan_quantum_2009, PhysRevA.94.063852, PhysRevE.88.012130, PhysRevE.87.012140, PhysRevLett.123.080602, PhysRevLett.127.190604,PhysRevLett.128.180602, PhysRevE.96.062120, bouton2021quantum,PhysRevResearch.5.043274, doi:10.1126/science.aad6320,gelbwaser2015work}. 

A quantum heat engine (QHE) is a quantum apparatus that performs a thermodynamic cycle from which one wishes to extract net work \cite{ 10.1116/5.0083192,zhang_quantum_2014, zhang_theory_2014,quan_quantum_2005, youssef_quantum_2010,serafini_optomechanical_2020, gluza_quantum_2021,PhysRevE.95.022135,niedenzu2018quantum,ghosh2019quantum,doi:10.1073/pnas.1110234108,doi:10.1126/science.1078955}. In quantum thermodynamics it is possible to define thermodynamic cycles, in complete analogy to the classical counterpart, that depend on the specific transformations performed by the system \cite{quan_quantum_2007,quan_quantum_2009}.  An example of a thermodynamic cycle largely considered in literature is the quantum Otto cycle \cite{quan_quantum_2007,e19040136,e20110875}, which consists of two adiabatic transformations and two isochoric transformations. At the quantum level, the isochoric transformation is equivalent to the thermalization of the working substance with the hot or cold bath, while the eigenenergies of the system remain constant. On the other hand, during each adiabatic transformation the quantum system is isolated, and its eigenenergies change by means of an external drive.
\\
The working substance releases net work during the adiabatic expansion. Nevertheless, the generation of work does not necessarily imply the release of power. This is the case when the adiabatic transformation occurs quasi-statically, for example. For technological purposes it is generally of greater interest to consider more practical scenarios, wherein transformations occur in finite time \cite{PhysRevResearch.2.033083,Alecce_2015,e22091060,PhysRevE.90.062134}, such that one can extract power from the working substance \cite{abah_single-ion_2012,rosnagel_nanoscale_2014, kosloff_quantum_2014, campisi_power_2016,
doi:10.1126/sciadv.adf1070}. In this case, it may happen that the release of power is affected by inner friction, which means that the system is still not exchanging heat with the environment (classical adiabatic condition), but the population varies during the transformation \cite{rezek_irreversible_2006,plastina2014irreversible}.

In this work we propose a quantum heat engine subject to inner friction that performs an Otto cycle whose adiabatic transformations mimic the outcome of the SU$(1,1)$ interferometer. The SU$(1,1)$ interferometer is a nonlinear interferometer that is constituted by active optical elements \cite{PhysRevA.33.4033,Seyfarth2020wignerfunction,10.1063/5.0004873}. In particular, it can be engineered from the standard Mach-Zehnder interferometer (MZI) by replacing the two beam splitters with two squeezers \cite{Chekhova:16}. The nonlinearity of the interferometer stems from the nonlinear susceptibility of the media typically used as squeezing resource \cite{Ferreri2021spectrallymultimode,sym14030552}.

In quantum metrology, there are several advantages of using SU$(1,1)$ interferometers with respect to other platforms, such as the Mach-Zehnder interferometer. For example, it has been shown that such devices are highly immune to external losses \cite{PhysRevA.86.023844}. Moreover, they can overcome the classical shot noise limit and reach the so-called Heisenberg limit even when seeded with the vacuum state \cite{PhysRevA.33.4033}.
For these reasons, this class of interferometers has been thoroughly studied in the past, for example by analyzing the role of different input states \cite{Adhikari:18,PhysRevA.95.063843,Plick_2010}, as well as specializing to both spectral and spatial multimode scenarios \cite{Frascella:19, Ferreri2021spectrallymultimode,sym14030552,PhysRevResearch.5.043158}.

The Otto cycle proposed in this work is based on two interacting quantum harmonic oscillators, whose time evolution is modelled in terms of elements of the $\mathfrak{su}(1,1)$ Lie algebra. In particular, our device exploits the equivalence between the output values of the observables after each adiabatic transformations and the outcomes at the end of an SU$(1,1)$ interference scheme to improve our knowledge about such observables beyond the shot-noise limit. 
Indeed, taken for granted that all observables of interest are affected by quantum fluctuations, we may ask if the uncertainty of our outcomes stems from the instability of the protocols controlling the adiabatic transformations, or from the quantum/thermal fluctuation of the observables themselves. To answer this question, we encode the information about our protocols into the phase of an equivalent SU$(1,1)$ interferometer and study the dependence of the phase sensitivity of the heat engine with respect to both the number of excitations and the average energy at the end of the adiabatic transformations, benchmarking it with the shot-noise limit. 

The phase sensitivity is a tool largely used in quantum metrology for estimating the minimal modulation of the phase needed to overcome the quantum (or in this case thermal) fluctuation of observables. In this work, the phase sensitivity is utilized as a tool to optimize the protocols such that, at the end of the adiabatic strokes, we can safely distinguish possible errors due to the instability of our protocols from the perturbation generated by the thermal fluctuations. We show that by increasing the amount of squeezing, while at the same time properly manipulating the internal phase, our device can work simultaneously as a quantum thermal machine performing the Otto cycle with a net work output, as well as an SU$(1,1)$ interferometer working beyond the shot-noise limit. 

The paper is structured as follows: in Section~\ref{tb} we review the mathematical tools to describe the two-mode SU$(1,1)$ interferometer as well as the concept of phase sensitivity. In Section~\ref{mhe} we introduce our model of SU$(1,1)$ heat engine, with focus on the analogy between the outputs of the adiabatic transformations and the outputs of an SU$(1,1)$ interference scenario. In Section~\ref{tr} we examine the performance of the QHE by analyzing both the efficiency of the cycle and the phase sensitivity at the end of the adiabatic expansion. We also discuss a superconducting platform where our model can be implemented. We report our conclusions in Section~\ref{cn}.
\section{Theoretical background}\label{tb}
In this section we introduce the formalism to study the system of interest. In particular, we first present the algebraic tools to describe the SU$(1,1)$ interferometer. Afterwards, we provide a short review of the concept of phase sensitivity and noise limits in quantum metrology.

\begin{figure}[t!]
	\centering
	\includegraphics[width=1\linewidth]{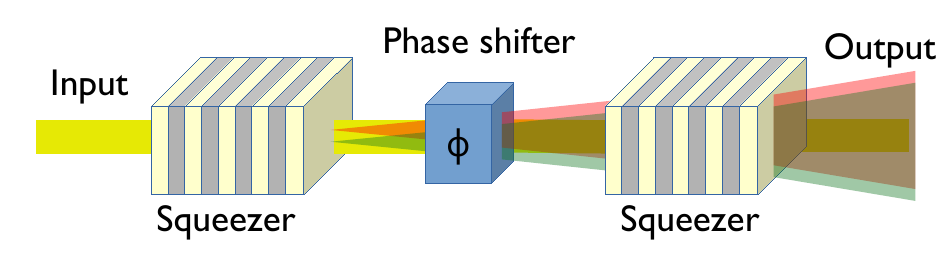}
	\caption{Design of the SU(1,1) interferometer. The input channel is seeded with a classical source (pump laser), which interacts with two squeezing sources (nonlinear optical media) in order to generate photon pairs. In the degenerate scheme, the created photons may be distinguished by their polarization. The two squeezers are separated by a phase shifter, which controls the relative phase between the laser and the photon pairs.}
 \label{su11fig}
\end{figure}

\subsection{The SU$(1,1)$ interferometer}
The SU$(1,1)$ interferometer is schematically depicted in Fig.~\ref{su11fig}, and it consists of a sequence of squeezing and phase operations: squeezing $\rightarrow$ phase shift $\rightarrow$ antisqueezing \cite{PhysRevA.33.4033}. The unitary transformation encoding the interference process takes the form
\begin{align}
\hat U_{\textrm{su}}(\zeta,\phi)=e^{-i\zeta \hat K_\textrm{x}}e^{-i\phi \hat K_\textrm{z}}e^{i\zeta \hat K_\textrm{x}},
\label{un1}
\end{align}
where $\zeta$ and $\phi$ identify the squeezing and the phase parameters, respectively.

The operators $K_i$ with $i=$x,y,z fulfill the following commutation relations, 
\begin{align}\label{su11:commutation:relations}
[\hat K_j,\hat K_k]= i f_{jk\ell} K_\ell,
\end{align}
which define the $\mathfrak{su}(1,1)$ Lie algebra with structure constants $f_{abc}$ that read: $f_{\textrm{xyz}}=-1$, $f_{\textrm{yzx}}=1$, $f_{\textrm{zxy}}=1$, while the only other non-vanishing expressions can be obtained by permutations of the indices in the standard fashion. The Lie algebra is fully defined by the commutation relations \eqref{su11:commutation:relations} and is independent of the choice of concrete representation of the operators $\hat{K}_j$. We also recall that the operators $\hat K_j$ satisfy the Jacobi identity $[A,[B,C]]+[B,[C,A]]+[C,[A,B]]=0$.

For the purposes of this work we choose to express these operators using the  annihilation and creation operators $\hat{a}_k,\hat{a}_k^\dag$ of two harmonic oscillators that satisfy the canonical commutation relations $[\hat{a}_k,\hat{a}_{k'}^\dag]=\delta_{kk'}$, while all others vanish. Employing these operators we introduce the following expressions 
\begin{align}
\hat K_\textrm{x}&=\frac{1}{2}(\hat a_1^\dag\hat a_2^\dag+\hat a_1\hat a_2), &
\hat K_\textrm{y}&=\frac{i}{2}(\hat a_1\hat a_2-\hat a_1^\dag\hat a_2^\dag),\nonumber\\
\hat K_\textrm{z}&=\frac{1}{2}(\hat a_1^\dag\hat a_1+\hat a_2\hat a_2^\dag), &
\hat N&=\hat a_1^\dag\hat a_1+\hat a_2^\dag\hat a_2,\label{Kz}
\end{align}
which include also the definition of the number operator $\hat{N}$ and we observe that $\hat{K}_\textrm{z}=(\hat N+1)/2$.
The Casimir invariant for this Lie algebra is $\hat K^2:=\hat K_\textrm{z}^2-\hat K_\textrm{x}^2-\hat K_\textrm{y}^2$, and it satisfies $[\hat{K}^2,\hat{K}_j]=0$ for all $j$.
Note that $\hat K_\textrm{z}$ is the only element of the group that commutes with the number operator $\hat N$. This feature is key, as we will see below.

An equivalent form of the unitary operator introduced in \eqref{un1} is given by
\begin{align}
\hat U_{\textrm{su}}=e^{i\theta \hat K_\textrm{z}}e^{i\chi\hat K_\textrm{y}}e^{-i\theta \hat K_\textrm{z}},
\label{un2}
\end{align}
where the parameters $\theta$ and $\chi$ can be determined as functions of $\zeta$ and $\phi$ by means of the following relations
\begin{align}
\cos(\theta(\zeta,\phi))=&\frac{\sin\phi}{\sqrt{\sin^2\phi+(1-\cos\phi)^2\cosh^2\zeta}},\nonumber\\
\cosh(\chi(\zeta,\phi))=&(1-\cos\phi)\cosh^2\zeta+\cos\phi,
\label{chi}
\end{align}
which have already been obtained in the literature \cite{PhysRevA.33.4033}.

The form of expression \eqref{un2} of the unitary operator $\hat U_{\textrm{su}}$ dramatically simplifies the computation of the average values of any observable that commutes with $\hat K_\textrm{z}$. To see this, consider a state diagonal in the Hamiltonian eigenbasis as initial state $\lvert \textrm{in}\rangle$ (as always will be in this work), and the final state $\lvert \textrm{out}\rangle$ at the end of the interference process. Then, we can easily compute the average number $N_\textrm{out}:=\langle \textrm{out}\rvert\hat N\lvert \textrm{out}\rangle$ of particles  leaving the interferometer, and we find 
\begin{align}
N_\textrm{out}=&
\langle\textrm{in}\rvert\hat U_{\textrm{su}}^\dag\hat N\hat U_{\textrm{su}}\lvert \textrm{in}\rangle\nonumber\\
=&\langle\textrm{in}\rvert e^{i\theta \hat K_\textrm{z}}e^{-i\chi\hat K_\textrm{y}}\hat N e^{i\chi\hat K_\textrm{y}}e^{-i\theta \hat K_\textrm{z}}\lvert \textrm{in}\rangle\nonumber\\
=& (N_\textrm{in}+1) \cosh{\chi}-1,
\end{align}
where $N_\textrm{in}:=\langle\textrm{in}\rvert\hat N\lvert \textrm{in}\rangle$ is the average number of particles entering the interferometer.
It is not surprising that the number of additional particles created is determined ultimately by the squeezing parameter $\chi$, since squeezing introduces energy in the system, and therefore potentially new excitations.

\subsection{Phase sensitivity}\label{ps}
Quantum metrology studies the precision that can be obtained by measurements of physical parameters when quantum resoruces can be exploited \cite{PhysRevD.23.1693, PhysRevD.26.1817}. Among possible applications one finds precision measurements using interferometers, where the performance of the interferometer can be evaluated by estimating its phase sensitivity \cite{demkowicz2015quantum}. This is the precision with which we can discriminate the variation of an observable $\hat O(\phi)$ due to the modulation of an internal parameter $\phi$ \cite{PhysRevA.93.023810}. At the output of the interferometer the average value $O(\phi):=\langle\hat O(\phi)\rangle_{\hat{\rho}}$ of the observable $\hat O(\phi)$ in the state $\hat{\rho}$ will depend on $\phi$. A small shift $\delta\phi$ of the variable $\phi$ induces a change $\delta O=O(\phi+\delta\phi)-O(\phi)=\frac{\partial O(\phi)}{\partial\phi}\delta\phi$, to first order in $\delta\phi$. To make sure that the variation of the observable is only due to the modulation of $\phi$, the variation itself must be at least as large as the statistical fluctuation of the observable itself, which translates in the condition $\delta O=\Delta O$, where $\Delta X$ determines the standard deviation of the quantity $X$. This means that 
\begin{align}
\delta\phi=\frac{\Delta O}{\left\lvert\frac{\partial O(\phi)}{\partial\phi}\right\rvert}
\label{phse}
\end{align}
is the variation of $\phi$ determining the smallest appreciable perturbation of $O$ beyond its statistical fluctuation. Note that an observable undergoing a Poissonian fluctuation achieves its best sensitivity at
\begin{align}
\Delta\phi_\textrm{SNL}=1/\sqrt{N_\phi},
\label{snl}
\end{align} 
which is called \textit{shot-noise limit} (SNL), and it is the maximum precision achieved by a Mach-Zehnder interferometer when the two input channels are seeded by a coherent state \cite{PhysRevD.26.1817}. In Eq.~\eqref{snl} the quantity $N_\phi$, indicates the number of photons undergoing the phase shift. A Mach-Zehnder interferometer overcoming this limit means therefore taking advantage of the sub-Poissonian (nonclassical) statistics of the input state to perform high precision measurement of $\delta O$. For instance, it was shown \cite{PhysRevD.23.1693, doi:10.1080/00107510802091298} that the phase sensitivity can scale as $\Delta\phi_\textrm{HL}=1/N_\phi$ if both input channels of the Mach-Zehnder interferometer are seeded with squeezed light. The scaling proportional to $N_\phi^{-1}$ is also called \textit{Heisenberg limit} (HL) since it is strictly connected to the energy-time uncertaintly principle \cite{PhysRevA.55.2598}. The crucial advantage of using the SU$(1,1)$ interferometer is the fact that it can overcome the SNL without requiring any exotic input state. In particular, it has been demonstrated that the precision of such interferometer can reach the Heisenberg limit by seeding it with the vacuum state \cite{PhysRevA.33.4033}.

\section{Model of a SU$(1,1)$ heat engine}\label{mhe}
\begin{figure}[t!]
	\centering
	\includegraphics[width=1\linewidth]{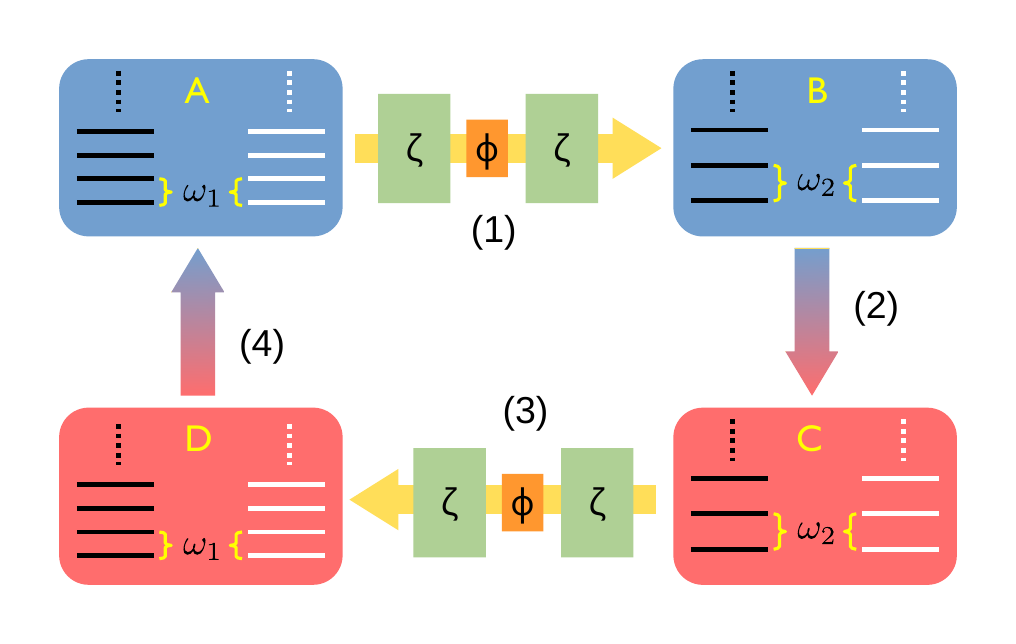}
	\caption{Pictorial representation of two degenerate quantum harmonic oscillators performing the SU$(1,1)$ Otto cycle. The four stages labeled by the capitol letters are connected by the four strokes of the cycle. In particular: (1) SU$(1,1)$ adiabatic compression ($A\rightarrow B$), (2) SU$(1,1)$ hot isochoric transformation ($B\rightarrow C$), (3) SU$(1,1)$ adiabatic expansion transformation ($C\rightarrow D$), and (4) cold isochoric ($D\rightarrow A$).}
 \label{ciclo}
\end{figure}

The system of interest is depicted in Fig.~\ref{ciclo}, and consists of two degenerate bosonic modes (i.e., with the same frequency), which are employed as working substance and perform an Otto cycle. 
The Otto cycle is a four strokes cycle in which the working substance undergoes four transformations:
\begin{enumerate}[label=(\arabic*)]
    \item \textit{Adiabatic compression}, during which the two oscillators, decoupled from any bath, increase their frequency; 
    \item \textit{Hot isocoric} thermalization of the system with a hot bath at temperature $T_\textrm{h}$ once the compression stops; 
    \item \textit{Adiabatic expansion}, during which the two oscillators release output work. Their frequency returns to the the same value that they had at the beginning of stage (1); 
\item \textit{Cold isochoric}, during which the system thermalizes with the cold bath at temperature $T_\textrm{c}$ and is ready to repeat the cycle. 
\end{enumerate}
Note that during each isochoric transformation the system is weakly coupled to the baths, which ensures that the dressing of the bare system Hamiltonian, due to the presence of nonvanishing system-bath interactions, can be ignored \cite{PhysRevE.94.012137}.

The frequency variation at each adiabatic transformation occurs by means of a unitary evolution operator $\hat U(t)$ defined in the next section. This operator encodes the information about the two protocols controlling both the time dependence of the bare frequency of the oscillators (which is identical) and the strength of their interaction. The unitary evolution operator varies according to the transformation it is referred to: during  compression, the frequencies are increased, whereas they decrease during the expansion. Crucially, we will see that the average values $O(t):=\langle \hat U^\dag(t)\hat O\hat U(t)\rangle_{\hat{\rho}}$ in the state $\hat{\rho}$ of any observable $\hat O$ of interest at the end of the time evolution are indistinguishable from its average value after the unitary transformation determined by Eq.~\eqref{un2}.

\subsection{Time evolution operator and Hamiltonian}\label{teo}
We assume that the adiabatic transformations causing the frequency shift of the two oscillators is represented by the following, already time-ordered, unitary operator:
\begin{align}
\hat U(t):=\overset{\leftarrow}{\mathcal{T}}e^{-i\int_0^t dt'\hat{H}_{\textrm{S}}(t')}=e^{-i f_\textrm{z}(t) \hat K_\textrm{z}}e^{-i f_\textrm{y}(t) \hat K_\textrm{y}},
\label{tiev}
\end{align}
where $f_\textrm{z}(t)$ and $f_\textrm{y}(t)$ are two time-dependent functions encoding the protocols of each adiabatic transformations. In our notation, we set $\hbar=1$.
Since the time evolution operator must be the identity operator $\mathds{1}$ at $t=0$, these two functions are subject to the initial conditions $f_\textrm{z}(0)=f_\textrm{y}(0)=0$.
The Hamiltonian written in the Schrödinger picture inducing such time evolution is obtained from Eq.~\eqref{tiev} by taking the time derivative on both sides and then multiplying each side by $\hat U^\dag(t)$ on the right \cite{bruschi_time_2013}. We find
{\small
\begin{align}
\hat{H}_{\textrm{S}}(t)=\dot{f}_\textrm{z}(t)\hat K_\textrm{z}+\dot{f}_\textrm{y}(t)[\cos(f_\textrm{z}(t)) \hat K_\textrm{y}-\sin(f_\textrm{z}(t))\hat K_\textrm{x}].
\label{hs}
\end{align}
}

We want the two oscillators to be decoupled at the beginning of each adiabatic transformation. In this way, the system can thermalize at each isochoric transformation without the two subsystems interacting. This means that, at $t=0$, we require $\dot f_\textrm{y}(0)=0$, and the initial Hamiltonian $\hat{H}_{\textrm{S}}(0)$ reduces to
\begin{align}
 \hat{H}_{\textrm{S}}(0)=\dot{f}_\textrm{z}(0)\hat K_\textrm{z}=\frac{\dot{f}_\textrm{z}(0)}{2}(\hat a_1^\dag\hat a_1+\hat a_2^\dag\hat a_2+1)   
\end{align}
This allows us to attribute the following initial conditions to the time derivative: $\dot{f}_\textrm{z}(0)=2\,\omega(0)=2\,\omega_{\textrm{i}}$, where $\omega_{\textrm{i}}$ is the frequency of the two oscillators before the time evolution. Importantly, at the end of the dynamics, that is at time  $t=t_{\textrm{f}}$, we re-initialize the Hamiltonian: thus, $\hat{H}_{\textrm{S}}(t_{\textrm{f}})=\hat{H}_{\textrm{S}}(0)$. This adds the following further boundary condition $\dot{f}_\textrm{y}(t_{\textrm{f}})=0$.

The Hamiltonian $\hat{H}_{\textrm{H}}(t)$ in the Heisenberg picture is obtained from Eq.~\eqref{tiev} via the relation 
\begin{align}
\hat{H}_{\textrm{H}}(t):=\hat U^\dag(t)\hat{H}_{\textrm{S}}(0)\hat U(t)\equiv -i\frac{d\hat U^\dag(t)}{dt}\hat U(t),
\end{align}
and it reads
{\small
\begin{align}
 \hat{H}_{\textrm{H}}(t)=&2\omega(t)[\text{ch}(f_\textrm{y}(t)) \hat K_\textrm{z}-\text{sh}(f_\textrm{y}(t)) \hat K_\textrm{x}]+\dot{f_\textrm{y}}(t)\hat K_\textrm{y},
\end{align}
}
where we identified $\dot{f}_\textrm{z}(t)=2\omega(t)$, and we have introduced ch$(x):=\cosh x$ and sh$(x):=\sinh x$ for simplicity of notation.

Note that if we had $\dot{f}_\textrm{y}(t)\approx 0$ at all times we would have $f_\textrm{y}(t)\approx 0$ due to the initial conditions, and the time evolution would describe a quantum adiabatic transformation \cite{quan_quantum_2007}. Once the dynamics stops at time $t_{\textrm{f}}$, the Hamiltonian $\hat{H}_{\textrm{H}}(t)$ becomes
\begin{align}
 \hat{H}_{\textrm{H}}(t_{\textrm{f}})&=2\omega_{\textrm{f}}[\text{ch}(f_\textrm{y}(t_{\textrm{f}})) \hat K_\textrm{z}-\text{sh}(f_\textrm{y}(t_{\textrm{f}})) \hat K_\textrm{x}],
 \label{hf}
\end{align}
where $\omega_{\textrm{f}}\equiv\omega(t_{\textrm{f}})$ is the frequency of the two oscillators at the end of the dynamics. The relation between the initial and final frequencies is $\omega_{\textrm{f}}>\omega_{\textrm{i}}$ at the end of the adiabatic compression, and $\omega_{\textrm{f}}<\omega_{\textrm{i}}$ at the end of the adiabatic expansion.

\subsection{The SU$(1,1)$ adiabatic transformation}\label{suad}
We now show that, although the unitary operator in Eq.~\eqref{tiev} driving the adiabatic transformation does not perfectly match the unitary operator of the SU$(1,1)$ interferometer in Eq.~\eqref{un2} (a second phase shift misses), the average values of any observable at the end of each adiabatic transformation are indistinguishable from the average values obtained as result of both the frequency tuning and the SU$(1, 1)$ interference of the bosonic modes.

Recall that the thermal state, which is the state of the two oscillators at the beginning of each adiabatic transformation, is a mixed state defined as $\rho=e^{-\beta\hat H_S(0)}/Z$, where $Z$ is the partition function and $\beta=1/(k_B T)$, with $T$ representing temperature and $k_B$ being the Boltzmann constant. In our case, since 
\begin{align}
\hat H_S(0)=-2\omega_{\textrm{i}}\hat K_\textrm{z}  
\end{align}
and
\begin{align}
\hat K_\textrm{z}=\hat N+1=\hat a_1^\dag\hat a_1+\hat a_2^\dag\hat a_2+1    
\end{align}
we have
\begin{align}
\hat{\rho}=&\sum_{n_1}\sum_{n_2}\frac{e^{-\omega_{\textrm{i}}\beta(n_1+n_2+1)}}{Z}\lvert n_1,n_2\rangle\langle n_1,n_2\rvert,
\label{thst}
\end{align}
where in the last line we expressed the thermal state in terms of the Hamiltonian (Fock) eigenstates. The partition function $Z:=\text{Tr}(\hat{\rho})$ therefore reads 
\begin{align}
Z=\sum_{n_1}\sum_{n_2}e^{-\omega_{\textrm{i}}\beta(n_1+n_2+1)}=\left[2\sinh\left(\beta\omega_{\textrm{i}}/2\right)\right]^{-2}.
\label{zeta}
\end{align}
Note that $\hat K_\textrm{z}$ and the Hamiltonian at $t=0$ share the same eigenbasis, because they commute. This means that, at the end of the interference process described by the unitary transformation in Eq.~\eqref{un2}, the average value $ O_\textrm{H}:=\langle \hat O_\textrm{H} \rangle_{\hat{\rho}} \equiv  \textrm{Tr}[\hat O_\textrm{H} \hat{\rho}]$ of any observable $\hat O_\textrm{H}$ calculated with respect to this thermal state must be of the form
\begin{align}
 O_\textrm{H}=&\textrm{Tr}[\hat U_\textrm{su}^\dag\hat O_\textrm{S}\hat U_\textrm{su} \hat{\rho}]\nonumber\\
=& \textrm{Tr}[e^{i\theta \hat K_\textrm{z}}e^{-i\chi\hat K_\textrm{y}}e^{-i\theta \hat K_\textrm{z}} \hat O_\textrm{S} e^{i\theta \hat K_\textrm{z}}e^{i\chi\hat K_\textrm{y}}e^{-i\theta \hat K_\textrm{z}} \hat{\rho}]\nonumber\\
=& \textrm{Tr}[e^{-i\chi\hat K_\textrm{y}}e^{-i\theta \hat K_\textrm{z}} \hat O_\textrm{S} e^{i\theta \hat K_\textrm{z}}e^{i\chi\hat K_\textrm{y}}\hat{\rho}],
\label{Oh}
\end{align}
where in the last step we took advantage of the commutation between $\hat K_\textrm{z}$ and $\hat{\rho}$, as well as the cyclic property of the trace.
The average value of $\hat O_\textrm{H}$ at the end of the interference process would therefore be indistinguishable from its average value at time $t_{\textrm{f}}$ if the time evolution is governed by the unitary operator in Eq.~\eqref{tiev}, assuming $\chi=-f_\textrm{y}(t_{\textrm{f}})$ and $\theta=-2\int_0^{t_{\textrm{f}}}dt'\omega(t')=-f_\textrm{z}(t_{\textrm{f}})$:
\begin{align}
 O_\textrm{H}=&\textrm{Tr}[\hat U_\textrm{su}^\dag\hat O_\textrm{S}\hat U_\textrm{su} \hat{\rho}]=\textrm{Tr}[\hat U^\dag(t)\hat O_\textrm{S}\hat U(t) \hat{\rho}].
 \end{align}
This proves that the average values of a observable calculated after the adiabatic transformation controlled via $\hat U(t)$ corresponds to the average value of the same observable when the system undergoes both the frequency tuning and the SU$(1,1)$ interference process.
As a consequence of these considerations we wish to strongly emphasize the fact that \textit{controlling the protocol functions at the end of the dynamics, $f_\textrm{y}(t_{\textrm{f}})$ and $f_\textrm{z}(t_{\textrm{f}})$, means controlling $\chi$ and $\theta$, which allows to directly manipulate both the squeezing parameter $\zeta$ and the phase $\phi$ of the equivalent SU$(1,1)$ interferometer by means of the constituent relations \eqref{chi}}.

In the study of the performance of the thermodynamic cycle presented here we note that the only observable of interest is the Hamiltonian, whose average value gives the amount of energy of the system at each step of the cycle. Employing Eqs.~\eqref{Kz}, \eqref{hf}, \eqref{thst}, \eqref{zeta}, and \eqref{Oh} we  finally obtain
\begin{align}
\langle\hat{H}_{\textrm{H}}(t_{\textrm{f}})\rangle_{\hat{\rho}}= \omega_{\textrm{f}}\cosh\chi\coth\left(\frac{\beta\omega_{\textrm{i}}}{2}\right).
\end{align}

\section{Analytical results}\label{tr}
We have introduced the working substance and the functioning of the SU$(1,1)$ adiabatic transformation. Therefore, we can now discuss the performance of the quantum heat engine. In particular, we will focus on the efficiency of the Otto cycle and the phase sensitivity of the equivalent SU$(1,1)$ interferometer at the end of the adiabatic expansion. Interestingly, we will see that it is possible to find a regime of parameters wherein our system \textit{can work simultaneously as a quantum heat engine producing net work, and as an SU$(1,1)$ interferometer with precision beyond the standard quantum limit.} Finally, we will discuss an application of our model of SU$(1,1)$ heat engine in the context of circuit QED~\cite{GU20171}.

\subsection{Thermodynamic cycle}
In this section we study the efficiency of the heat engine. Therefore, we start from calculating the average value of the energy at each of the four stages of the cycle. To do this, we calculate the average value of the Hamiltonian at the end of the two thermalization branches (isochoric transformation) and at the end of the two adiabatics:
\begin{align}
\langle \hat{H}\rangle_A&= \omega_1\coth\left(\frac{\beta_\textrm{c}\omega_1}{2}\right),\\
\langle \hat{H}\rangle_B&= \omega_2\cosh\chi\coth\left(\frac{\beta_\textrm{c}\omega_1}{2}\right),\\
\langle \hat{H}\rangle_C&= \omega_2\coth\left(\frac{\beta_\textrm{h}\omega_2}{2}\right),\\
\langle \hat{H}\rangle_D&= \omega_1\cosh\chi\coth\left(\frac{\beta_\textrm{h}\omega_2}{2}\right),
\end{align}
where $\omega_1$ and $\omega_2$ are the oscillators frequency at the beginning and at the end of the compression, respectively, whereas $\beta_\textrm{h}=1/(k_\textrm{B}T_\textrm{h})$ and $\beta_\textrm{c}=1/(k_\textrm{B}T_\textrm{c})$ are the inverse temperatures of the hot and cold  baths, respectively. Here $k_\textrm{B}$ is the Boltzmann constant as usual. 

From these expressions, we can calculate the average work and heat transferred during each transformation. During the adiabatic compression, the external work done on the system is given by $W_{AB} := \langle \hat{H} \rangle_B - \langle \hat{H} \rangle_A$, which becomes:
\begin{align}
\frac{W_{AB}}{\omega_1}&=\left(\frac{\omega_2}{\omega_1}\cosh\chi-1\right)\coth\left(\frac{\beta_\textrm{c}\omega_1}{2}\right).
\label{wab}
\end{align}
The system is then thermalized to the hot bath absorbing the amount $Q_{BC}=\langle \hat{H}\rangle_C-\langle \hat{H}\rangle_B$ of heat from the environment, given by
\begin{align}
\frac{Q_{BC}}{\omega_2}&=\coth\left(\frac{\beta_\textrm{h}\omega_2}{2}\right)-\cosh\chi\coth\left(\frac{\beta_\textrm{c}\omega_1}{2}\right).
\label{qbc}
\end{align}
At this point, the system releases energy in the form of fruitful work defined by $W_{CD}=\langle \hat{H}\rangle_D-\langle \hat{H}\rangle_C$. We find
\begin{align}
\frac{W_{CD}}{\omega_2}=\left(\frac{\omega_1}{\omega_2}\cosh\chi-1\right)\coth\left(\frac{\beta_\textrm{h}\omega_2}{2}\right).
\label{wreal}
\end{align}
Notice that $W_{AB}$ can be obtained from $W_{CD}$ by exchanging the two frequencies (and vice versa), as well as the hot and cold temperatures.

Finally, the system thermalizes with the cold bath ceding heat to the environment defined by 
$Q_{DA}:=\langle \hat{H}\rangle_A-\langle \hat{H}\rangle_D$, which reads
\begin{align}
\frac{Q_{DA}}{\omega_1}=\coth\left(\frac{\beta_\textrm{c}\omega_1}{2}\right)-\cosh\chi\coth\left(\frac{\beta_\textrm{h}\omega_2}{2}\right).
\end{align}
The cycle can therefore restart from the compression stage $A\rightarrow B$.

The explicit expressions obtained above allow us to calculate the efficiency of the cycle. This is defined as the ratio between the net work and the heat absorbed by the system. We employ Eqs.~\eqref{wab}, \eqref{qbc} and \eqref{wreal} to obtain $\eta=-(W_{AB}+W_{CD})/{Q_{BC}}$, which for us reads
\begin{align}
\eta&=1-\frac{\omega_1}{\omega_2}\frac{\cosh\chi\coth\left(\frac{\beta_\textrm{h}\omega_2}{2}\right)-\coth\left(\frac{\beta_\textrm{c}\omega_1}{2}\right)}{\coth\left(\frac{\beta_\textrm{h}\omega_2}{2}\right)-\cosh\chi\coth\left(\frac{\beta_\textrm{c}\omega_1}{2}\right)}
\label{eff}
\end{align}

This result, unsurprisingly, is very similar to what achieved in the literature \cite{abah_single-ion_2012}, where the coefficients $Q_1^*$ and $Q_2^*$, in the same manner of our parameter $\chi$, encode the adiabaticity of the compression and the expansion, respectively. Consequently, an analysis of the output power (which also includes the study of the efficiency at maximum power) would not substantially differ from what accomplished before \cite{abah_single-ion_2012}, and therefore we choose not to report it here.

The squeezing during each adiabatic process causes the presence of quantum friction. Specifically, during the expansion, the amount of uncontrollable work can be computed as described in \cite{plastina2014irreversible}. In our case, it is given by $W_\textrm{fric} = W_{CD} - W_\textrm{ad}$, which explicitly reads
\begin{align}
W_\textrm{fric}=&2\omega_1\sinh^2(\chi/2)\coth\left(\frac{\beta_\textrm{h}\omega_2}{2}\right).
\end{align}
Here, the work exchanged during the expansion, assuming the process were quantum adiabatic, is given by \begin{align}
W_\textrm{ad}=\left(\omega_1-\omega_2\right)\coth\left(\frac{\beta_\textrm{h}\omega_2}{2}\right).
\end{align}

Evidently, the higher the parameter $\chi$, the smaller the amount of energy extracted from the system during the expansion. In general, in order for our system to work as quantum heat engine we require $-(W_{AB}+W_{CD})>0$, or $\chi<\chi_\textrm{max}$, where we have defined $\chi_\textrm{max}$ through
\begin{align}
\cosh(\chi_\textrm{max})<\frac{\omega_1\coth\left(\frac{\beta_\textrm{c}\omega_1}{2}\right)+\omega_2\coth\left(\frac{\beta_\textrm{h}\omega_2}{2}\right)}{\omega_2\coth\left(\frac{\beta_\textrm{c}\omega_1}{2}\right)+\omega_1\coth\left(\frac{\beta_\textrm{h}\omega_2}{2}\right)}.
\label{chimax}
\end{align}
Interestingly, if $k_\textrm{B}T_\textrm{h}\ll \hbar \omega_2$ and $k_\textrm{B}T_\textrm{c}\ll \hbar \omega_1$, we have $\coth\left(\beta_\textrm{c}\omega_1/2\right)\approx 2T_\textrm{c}/\omega_1$ and $\coth\left(\beta_\textrm{h}\omega_2/2\right)\approx 2T_\textrm{h}/\omega_2$, and we can express the positive work condition in terms of the temperature ratio as
\begin{align}
\frac{T_\textrm{h}}{T_\textrm{c}}>\frac{\omega_2}{\omega_1}\frac{\omega_2 \cosh\chi-\omega_1}{\omega_2-\omega_1\cosh\chi}.
\end{align}
As expected, the right side of the equation above reduces to the frequency ratio $\omega_2/\omega_1$ in case of quantum adiabatic transformations, $\chi=0$.

Note that Eq.~\eqref{chimax} corresponds to an upper bound for the parameter $\chi$. Nevertheless, this bound does not necessarily affect our choice of the squeezing parameter $\zeta$. From Eq.~\eqref{chi}, we see that we can choose relatively large $\zeta$ at the price of a narrower range for $\phi$. With $\zeta$ fixed, we can easily find that the phase takes values in the interval $0\leq\phi<\phi_\textrm{max}$, where $\phi_\textrm{max}$ is defined as:
\begin{align}
\cos(\phi_\textrm{max}(\zeta))=1-2\frac{\sinh^2(\chi_\textrm{max}/2)}{\sinh^2\zeta}.
\label{phimax}
\end{align}
Notably, the higher $\zeta$ the more $\phi_\textrm{max}$ tends to vanish.

\subsection{Efficiency at minimum sensitivity}

\begin{figure*}[t!]
	\centering
	\includegraphics[width=1\linewidth]{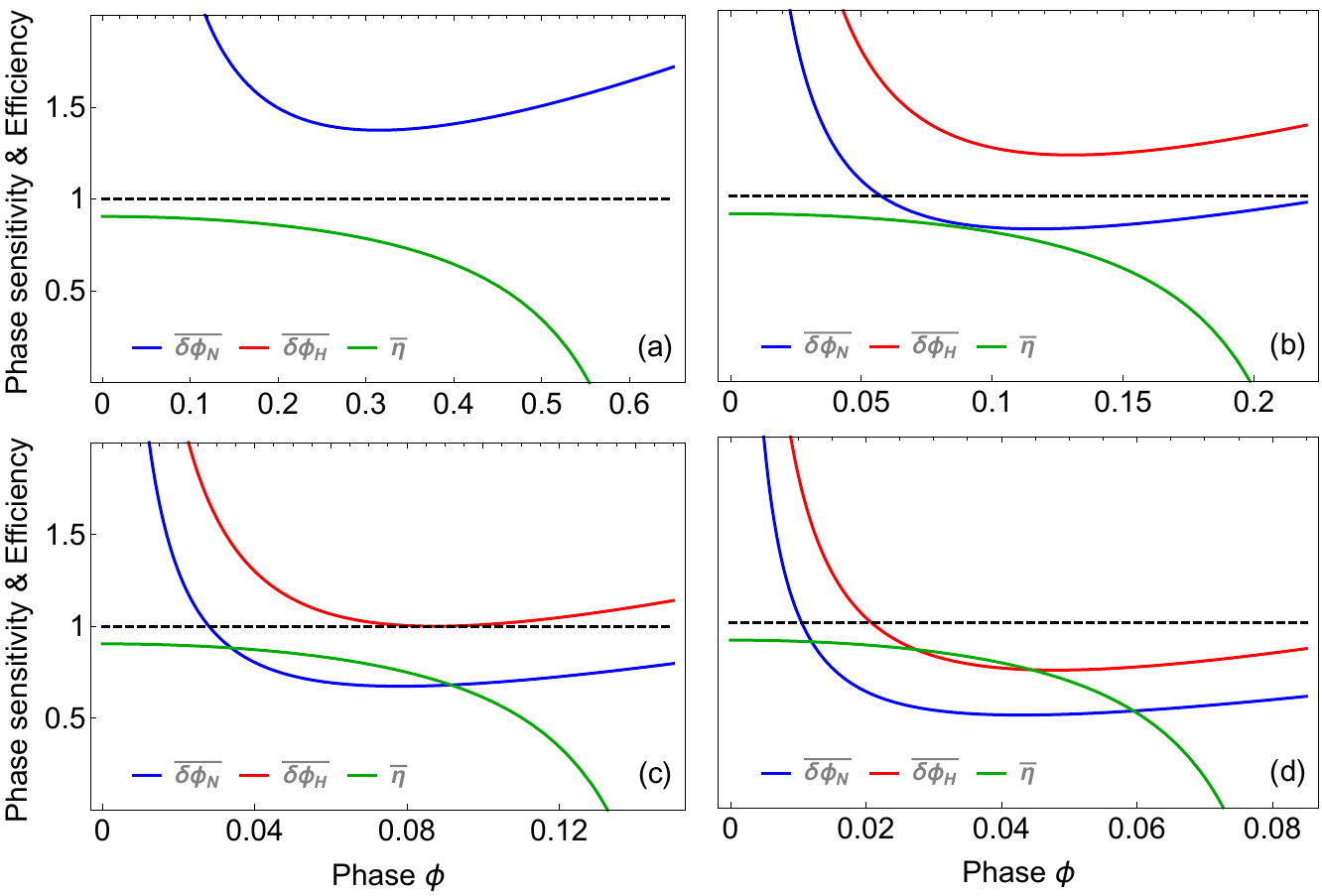}
	\caption{Normalized phase sensitivity calculated with respect to the average excitation number (blue curve) and the average energy (red curve), normalized efficiency of the Otto cycle (magenta curve). The bar above all quantities in the legend stresses these normalization. The dashed black horizontal line indicates the normalization. The four panels refer to different squeezing intensities: $\zeta=2$ (a), $\zeta=3$ (b), $\zeta=3.4$ (c), and $\zeta=4$ (d).
 Other parameters are: $\omega_1=0.1$, $\omega_2=1$, $T_\textrm{h}=2$ and $T_\textrm{c}=0.01$. Frequencies and temperatures are normalized with respect to $\omega_2$.}
 \label{graph}
\end{figure*}
It should be clear from Section~\ref{suad} that we can interpret the final stage of adiabatics of the quantum heat engine as an interference process. The parameters $\theta$ and $\chi$, representing the protocols of the time evolution operators at $t_{\textrm{f}}$, can be parametrized in terms of the squeezing parameter $\zeta$ and the internal phase $\phi$ of a SU$(1,1)$ interferometer by means of Eq.~\eqref{chi}. 

Assuming that the adiabatic processes induce an amount of squeezing quantified by $\zeta$, and that the source of instability in our protocols is entirely captured by the phase $\phi$, we can then ask what precision can we achieve from the knowledge of an observable parametrically dependent on $\phi$ and subject to thermal noise. 

In other words, since $\theta$ and $\chi$ are connected to the protocols of the QHE at the end of the adiabatics, we may ask how precisely we can attribute the uncertainty of an observable to the instability of $f_\textrm{z}(t_{\textrm{f}})$ and $f_\textrm{y}(t_{\textrm{f}})$, rather than to its own thermal fluctuation. 

Using the mathematical tools introduced in Section~\ref{ps}, we can make use of the phase sensitivity to test the precision of our knowledge of relevant observables, such as the output number of particles or, more importantly, the average energy at the end of the process. 

Note that, although the study of  work fluctuations has a fundamental role in quantum thermodynamics \cite{talkner_fluctuation_2009, campisi_fluctuation_2009}, work is not an observable \cite{PhysRevE.75.050102,campisi_colloquium_2011}, but it is calculated as the difference between the average energy of the working substance after the adiabatic transformation (which depends on the internal phase of the interferometer) and the average energy before the adiabatic transformation (which does not depend on the phase $\phi$). For these reasons, we discuss only those observables which are directly involved by the variation of $\phi$ in our study of the phase sensitivity.

According to the definition in Eq.~\eqref{phse}, we need to calculate both the variance of our observables and their derivative with respect to $\varphi$. Focusing on the adiabatic expansion (which is more affected by thermal fluctuations), we compute $\Delta^2 N=\textrm{Tr}[N_\textrm{H}^2\rho]-\textrm{Tr}[N_\textrm{H}\rho]^2$ and $\Delta^2 H=\textrm{Tr}[H_\textrm{H}^2\rho]-\textrm{Tr}[H_\textrm{H}\rho]^2$, which read
\begin{align}
\Delta^2 N&=\frac{1}{2}\left[\cosh(2\chi)\coth^2\left(\frac{\beta_\textrm{h}\omega_2}{2}\right)-1\right],\\
\Delta^2 H&=2\omega_1^2\left[\Delta^2 N+\frac{1}{4}\left(\coth^2\left(\frac{\beta_\textrm{h}\omega_2}{2}\right)+1\right)\right],
\end{align}

whereas the latter are
\begin{align}
\frac{\partial N_\textrm{H}}{\partial\phi}=&\sin\phi\sinh^2\zeta\coth^2\left(\frac{\beta_\textrm{h}\omega_2}{2}\right),\\
\frac{\partial H_\textrm{H}}{\partial\phi}=&\omega_1\frac{\partial N_\textrm{H}}{\partial\phi}.
\end{align}
Clearly, both variance and derivative depend on the phase of the interferometer $\phi$. We notice that the variance of the energy is not proportional to $\Delta N$, and this is due to the fact that the Hamiltonian operator also includes the energy of the quantum vacuum.

The two phase sensitivity curves are plotted in Fig.~\ref{graph}, along with the efficiency of the thermodynamic cycle in Eq.~\eqref{eff}, with respect to the phase $\phi$. Here, the indices ``$N$" and ``$H$" indicate the phase sensitivity calculated with respect to the excitation number and the Hamiltonian, respectively. We normalized the curves with respect to natural benchmarks: the phase sensitivity is normalized with respect to the SNL, while the efficiency is normalized with respect to the Carnot limit $\eta_\textrm{C}=1-T_\textrm{c}/{T_\textrm{h}}$.

Recall that the SU$(1,1)$ interferometer consists of active elements that do not preserve the number of particles. For this reason, in the definition of SNL in Eq.~\eqref{snl} we need the actual number of particles subject to the phase modulation. We therefore have $\Delta\phi_\textrm{SNL}=1/\sqrt{N_\phi}$, where
\begin{align}
N_\phi=(N_\textrm{in}+1) \cosh{\zeta}-1.
\end{align}

At $\phi=0$ we have $\chi=0$, and the quantum system performs quantum adiabatic transformations reaching therefore the efficiency of an ideal Otto cycle $\eta_\textrm{O}=1-\omega_1/\omega_2$ \cite{quan_quantum_2007}. As soon as we increase the phase $\phi$, the efficiency inevitably decreases due to the inner friction until it vanishes at $\phi_\textrm{max}$.
Note that, as predicted in Eq.~\eqref{phimax}, the operational phase range of our quantum system as a heat engine decreases with increasing squeezing $\zeta$.

We now look at the two phase sensitivity curves. We observe that, as the squeezing parameter $\zeta$ increases, the phase sensitivity decreases. For very large values of $\zeta$, both curves reach their minimum values below the SNL line (indicated by the black dotted line in the graph). In this regime, our quantum system functions both as a quantum heat engine and as SU$(1,1)$ interferometer working beyond the classical limit. 

In quantum metrology, the range of $\phi$ where the phase sensitivity overcomes the SNL, i.e., $\delta\phi<\Delta\phi_\textrm{SNL}$, is sometimes called supersensitivity \cite{PhysRevLett.119.223604, Ferreri2021spectrallymultimode, PhysRevResearch.5.043158}. Evidently, the supersensitivity range depends on the observable with respect to which the phase sensitivity is calculated: from Fig.~\ref{graph}b, c and d, we see that, for a fixed squeezing parameter $\zeta$, $\delta\phi_N$ is always lower than $\delta\phi_H$, and the corresponding supersensitivity range is larger.

At this point, we may be interested at the efficiency of the cycle when the minimum of the phase sensitivity $\Delta\phi_H$ reaches the SNL. To calculate this quantity, we first need to calculate the minimum $\zeta$ required to reach the SNL.
Given a set of $\{\omega_1, \omega_2, T_\textrm{h}, T_\textrm{c}\}$ of frequencies and temperatures, this value determines the minimum amount of squeezing necessary for the phase sensitivity to reach the SNL in at least one point of the phase $\phi$, which we will call $\phi_\textrm{SNL}$:
\begin{align}
\delta\phi_\textrm{min}(\zeta_\textrm{SNL})=\Delta\phi_\textrm{SNL}
\label{psmin}
\end{align}
The corresponding phase $\phi_\textrm{SNL}$ is obtained from Eq.~\eqref{psmin} and used to calculate the efficiency at the SNL.
This reads
\begin{align}
\eta&=1-\frac{\omega_1}{\omega_2}\frac{\cosh\chi_\textrm{SNL}\coth\left(\frac{\beta_\textrm{h}\omega_2}{2}\right)-\coth\left(\frac{\beta_\textrm{c}\omega_1}{2}\right)}{\coth\left(\frac{\beta_\textrm{h}\omega_2}{2}\right)-\cosh\chi_\textrm{SNL}\coth\left(\frac{\beta_\textrm{c}\omega_1}{2}\right)}
\label{eff}
\end{align}
where $\chi_\textrm{SNL}\equiv\chi(\zeta_\textrm{SNL}, \phi_\textrm{SNL})$.  These quantities are not simple to calculate analytically. However, we numerically computed both $\zeta_\textrm{SNL}$ and $\eta_\textrm{SNL}$ with the choice of parameters used in Fig.~\ref{graph}. This result in $\zeta_\textrm{SNL}=3.4$ and $\eta_\textrm{SNL}\simeq 0.705$. For completeness, we also report the efficiency at $\phi=0$, $\eta_\textrm{O}=0.9$, and the Carnot efficiency, $\eta_\textrm{C}=0.995$.

\subsection{Application to superconducting transmission lines}
The specific dynamics described by the unitary time evolution operator in Eq.~\eqref{tiev} can be achieved in circuit QED~\cite{RevModPhys.84.1, blais_circuit_2021}.
Thanks to their flexibility and high controllability, superconducting circuits are indeed promising platforms for the realization of QHE~\cite{quan_maxwells_2006,pekola_towards_2015,karimi_otto_2016,pekola_thermodynamics_2019,PhysRevB.86.014501,PhysRevA.85.063811,PhysRevLett.101.253602,PhysRevApplied.17.064022}. Moreover, transmission lines simulating particle creation phenomena~\cite{RevModPhys.84.1}, such as the dynamical Casimir effect~\cite{PhysRevLett.103.147003,PhysRevA.82.052509,wilsonObservationDynamicalCasimir2011}, have already been taken into account for the study of effects of quantum friction on the efficiency of the Otto cycle~\cite{del_grosso_quantum_2022}.

\begin{figure}[t!]
	\centering
	\includegraphics[width=1\linewidth]{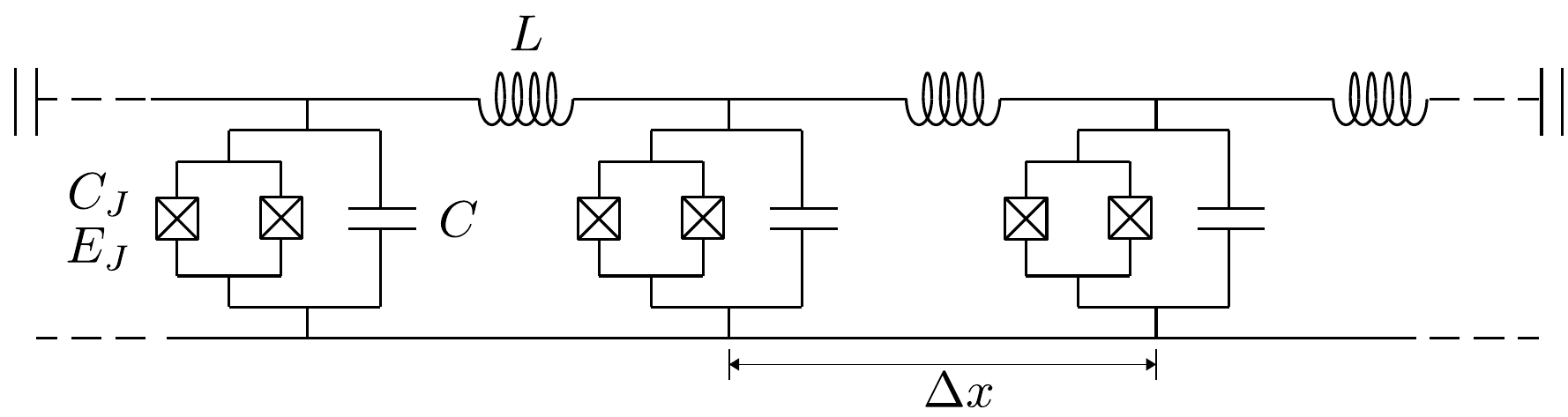}
	\caption{Example of platform supporting the SU$(1,1)$ Otto cycle. The superconducting transmission line consists of a series of inductors, while two consecutive inductors are connected by a node with a SQUID and a capacitor, placed in parallel.}
 \label{circ}
\end{figure}

Here we consider the device drawn in Fig.~\ref{circ}, and discussed in detail in~\cite{tian_analog_2017,tian_analogue_2019}, as possible platform for the implementation of the SU$(1,1)$ Otto cycle proposed here. The transmission line is based on a series of unit cells. Each cell consists of a inductor placed in series, and a superconducting quantum interference device (SQUID) placed in parallel. Finally, each SQUID is parallel to a capacitor.

This transmission line has been proposed as platform to simulate the cosmological model of particle creation due to a rapid expansion of the spacetime~\cite{tian_analog_2017,bernard1977regularization,birrell1984quantum}. The Lagrangian, as well as the equations of motion, describing the quantum magnetic flux field along the transmission line can indeed be interpreted as a (1+1)-dimensional equivalent of the Lagrangian of a quantum massive scalar field immersed in a dynamical background described by a time-dependent Friedmann-Robertson-Walker metric~\cite{bernard1977regularization,birrell1984quantum}. It emerges that the modes of the quantum magnetic flux undergo the same Bogoliubov transformations leading to the squeezing of the quantum vacuum of the scalar field before the expansion process. 

For later convenience, we report here the dispersion relation of the transmission line, as well as the Bogoliubov transformations. The former has already been obtained \cite{tian_analog_2017, PhysRevResearch.6.033204}, and reads
\begin{align}
\omega_j(t)=\sqrt{\frac{4\sin^2\left(\frac{k_j\Delta x}{2}\right)}{L C}+\left(\frac{2\pi}{\Phi_0}\right)^2\frac{E(t)}{C}},
\end{align}
where $k_j=2\pi j/(N_\textrm{cell}\Delta x)$ are the wave vectors of the quantum field assuming to possess a transmission line with $N_\textrm{cell}$ cells, $\Delta x$ is the cell length, $L$ and $C$ are respectively the inductance and the effective capacitance of the transmission line, $\Phi_0$ is the magnetic flux quantum, and $E(t)=E_0 [A\pm B\tanh(\nu\, t)]$ is the time-dependent Josephson energy, with $A$ and $B$ adimensional constants and $\nu$ rapidity coefficient. The sign in the Josephson energy depends on the thermodynamic transformation we are considering: the positive sign refers to the frequency during the adiabatic compression, while the negative sign refers to the expansion \cite{note}.
The Bogoliubov transformations have the general expression
\begin{align}
\hat a_j^{\textrm{out}}=\alpha_j \hat a_j^\textrm{in}+\beta_j^* (\hat a_{-j}^\textrm{in})^\dag, 
\label{bogtr}
\end{align}
which is well known in the literature \cite{bernard1977regularization, birrell1984quantum, tian_analog_2017}, while the Bogoliubov coefficients $\alpha_j,\beta_j$ for this particular case read
\begin{align}
\alpha_j=&\left(\frac{\omega_{\textrm{f}}}{\omega_{\textrm{i}}}\right)^\frac{1}{2}\frac{\Gamma(1-i\omega_{\textrm{i}}/\nu)\Gamma(-i\omega_{\textrm{f}}/\nu)}{\Gamma(-i\omega_+/\nu)\Gamma(1-i\omega_+/\nu)},\nonumber\\
\beta_j=&\left(\frac{\omega_{\textrm{f}}}{\omega_{\textrm{i}}}\right)^\frac{1}{2}\frac{\Gamma(1-i\omega_{\textrm{i}}/\nu)\Gamma(i\omega_{\textrm{f}}/\nu)}{\Gamma(i\omega_-/\nu)\Gamma(1+i\omega_-/\nu)}.
\end{align}
Here, we define $\omega_{\textrm{i}}\equiv \omega_{j}(t\rightarrow-\infty)$ and $\omega_{\textrm{f}}\equiv \omega_{j}(t\rightarrow+\infty)$  as the frequencies of the degenerate modes before and after the transformation, respectively. Additionally, $\omega_{+}=\omega_{\textrm{i}}+\omega_{\textrm{f}}$ and $\omega_{-}=\omega_{\textrm{f}}-\omega_{\textrm{i}}$.

We can safely isolate any degenerate mode pair of the transmission line by properly fixing the boundary conditions of the magnetic flux field. Therefore, we can omit the index $j$, focus our attention on the first mode pair, and write the Hamiltonian at the beginning of the dynamics as
\begin{align}
    \hat H_\textrm{S}=\omega_{\textrm{i}}[(\hat a_1^\textrm{in})^\dag\hat a_1^\textrm{in}+(\hat a_2^\textrm{in})^\dag\hat a_2^\textrm{in}+1].
    \label{Hs1}
\end{align}
Note that the two modes are distinguishable, thus \textit{degenerate}. At the end of the dynamics, the Hamiltonian in the Heisenberg picture is 
\begin{align}
\hat H_\textrm{H}\equiv\omega_{\textrm{f}}\left[(\hat a_1^\textrm{out})^\dag\hat a_1^\textrm{out}+(\hat a_2^\textrm{out})^\dag\hat a_2^\textrm{out}+1\right],  
\end{align}
which reads
\begin{align}
\hat H_\textrm{H}
=&\left(1+2\lvert\beta\rvert^2\right)\left[(\hat a_1^\textrm{in})^\dag\hat a_1^\textrm{in}+(\hat a_2^\textrm{in})^\dag\hat a_2^\textrm{in}+1\right]\nonumber\\
&+2\Re\{\alpha\beta\}\left[\hat a_1^\textrm{in}\hat a_2^\textrm{in}+(\hat a_1^\textrm{in})^\dag(\hat a_2^\textrm{in})^\dag\right]\nonumber\\
&+2i \Im\{\alpha\beta\}\left[\hat a_1^\textrm{in}\hat a_2^\textrm{in}-(\hat a_1^\textrm{in})^\dag(\hat a_2^\textrm{in})^\dag\right]
\label{Hab}
\end{align}
in terms of the initial operators, and we have used the Bogoliubov identity $|\alpha|^2-|\beta|^2=1$ for this specific one-dimensional degenerate case.
\begin{figure}[t!]
	\centering
	\includegraphics[width=1\linewidth]{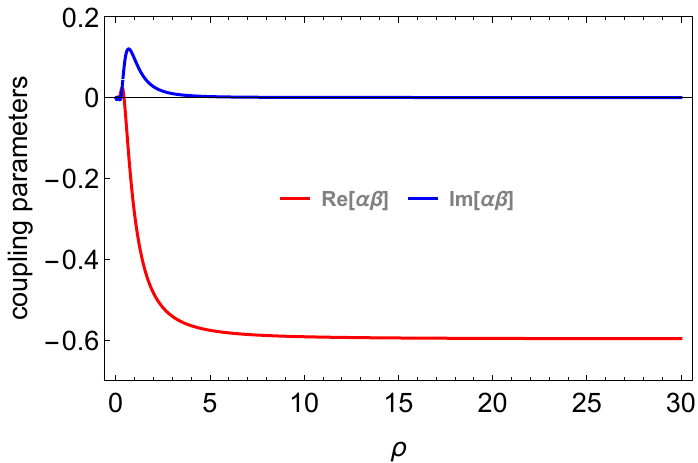}
	\caption{Dependence of the coupling coefficients $\Re\{\alpha\beta\}$ (red curve) and $\Im\{\alpha\beta\}$ (blue curve) on the parameter $\nu$. When $\nu$ is much larger than the two frequencies $\omega_{\textrm{i}}$ and $\omega_{\textrm{f}}$, the imaginary part of $\alpha\beta$  vanishes. Chosen values: $\omega_{\textrm{i}}=1$ and $\omega_{\textrm{f}}=0.35$. Frequencies and temperatures are normalized with respect to $\omega_{\textrm{i}}$.}
 \label{graph2}
\end{figure}

Interestingly for our goals, it turns out that the Bogoliubov transformations in Eq.~\eqref{bogtr} fulfill the following properties:
\begin{itemize}
\item[(i)] they describe a two-mode squeezing scenario;
\item[(ii)] they couple two degenerate modes  (in the specific case, counterpropagating modes with the same frequency and opposite momentum);
\item[(iii)] we can find a regime wherein the Bogoliubov coefficients can take real values.
\end{itemize}
The first two conditions are evident from Eqs.~\eqref{bogtr} and \eqref{Hs1}. However, in order to demonstrate that there is a regime wherein condition (iii) applies (i.e., the last term of Eq.~\eqref{Hab} vanishes), in Fig.~\ref{graph2} we plotted the two coupling parameters, namely the coefficients in the last two lines of Eq.~\eqref{Hab}, by fixing two convenient values for $\omega_{\textrm{i}}$ and $\omega_{\textrm{f}}$ and varying the rapidity parameter $\nu$. The graph shows that, when $\nu$ is much larger than the frequencies of the system, the imaginary part of the coupling constant vanishes, and the Hamiltonian in Eq.~\eqref{Hab} corresponds to that in Eq.~\eqref{hf}, with $\cosh(f_\textrm{y}(t_\textrm{f}))=1+2\lvert\beta\rvert^2$ and $\sinh(f_\textrm{y}(t_\textrm{f}))=-2\Re\{\alpha\beta\}$.

Despite the strong analogy between the dynamics in the transmission line and the adiabatic transformation described in this work, we need to clarify an important point.
Our model interprets the adiabatic transformation of the Otto cycle as the result of a time evolution that starts at $t=0$ and terminates at $t=t_\textrm{f}$. Once the dynamics stops, we control all relevant coefficients at time $t=t_\textrm{f}$ in order to optimize the phase sensitivity. Instead, the frequency transformation in the transmission line described by the Bogoliubov transformation in Eq.~\eqref{bogtr} refers to an interaction that occurs in an infinite time, from $t=-\infty$ to $t=+\infty$. The two values of the frequency, $\omega_{\textrm{i}}$ and $\omega_{\textrm{f}}$, must therefore be intended as the frequencies of the two oscillators in the past and in the future, respectively, as defined below Eq.~\eqref{bogtr}. Nevertheless, if the transition between the two values happens in a short time (or alternatively, if $\nu\gg\omega_{\textrm{i}},\omega_{\textrm{f}}$), we can make the reasonable assumption that the adiabatic transformation occurs in a finite time, and that the frequency of the two oscillator in the past and in the future are respectively the effective initial and final frequency of the two oscillators. Moreover, at $t_{\textrm{f}}\gg 1/\nu$, namely at time $t$ far from the frequency transition, we can safely assume that $\omega_{\textrm{f}}$ is constant. This means that the parameter $\theta$ approximately takes the value $\theta\approx-\omega_{\textrm{f}}\,t_\textrm{f}$. 

\begin{table}[h]
\centering
\caption{Choice of parameters}
\begin{tabular}{lll}
\hline
\hline
$T_\textrm{h}$ &\qquad 2 K \\
$T_\textrm{c}$ &\qquad 0.01 K \\
$C$ &\qquad 0.4 pF \\
$L$ &\qquad 60 pH \\
$E_0/C$ &\qquad 1 nJ$\times\,\textrm{F}^{-1}$ \\
$N_\textrm{cell}$ 	&\qquad 100\\
$A$ &\qquad 1 \\
$B$ &\qquad 0.78 \\
$\nu$ &\qquad 20
\end{tabular}
  \label{tab:overview}
\end{table}  

Recalling that $\chi\equiv -f_\textrm{y}(t_\textrm{f})$, and that we can control $\phi$ and $\zeta$ by means of $\theta$ and $\chi$ [see Eqs.~\eqref{chi}], we can now implement the SU$(1,1)$ Otto cycle in the superconducting circuit. To realize the two isochoric transformations, we need to couple the transmission line with the two baths, whereas the two adiabatic transformations are accomplished by tuning the Josephson energy.
We observe that, with an appropriate choice of circuit parameters, we can reach a regime wherein the transmission line behaves both as Otto heat engine and as supersensitive SU$(1,1)$ interferometer during each adiabatic transformation. In particular, we obtain the normalized efficiency $\overline\eta=0.23$ and the normalized phase sensitivity $\overline{\Delta\phi}=0.56$ with the choice of parameters reported in Table~\ref{tab:overview}.

\subsection{Why the SU(1,1) interferometer?}
We believe it is necessary to stress the fundamental role of the $\mathfrak{su}(1,1)$ Lie algebra employed for the description of the SU(1,1) interferometer. One may wonder for example if it is possible to realize a similar model of QHE based on other type of interferometers, for instance on the well-known Mach-Zehnder interferometer. The algebra describing the transformations of the MZI is the $\mathfrak{su}(2)$ Lie algebra, which can be used to describe the rotations in the three-dimensional space \cite{PhysRevA.33.4033}. Choosing this interferometer would however turn out to be inconvenient: all elements of the $\mathfrak{su}(2)$ Lie algebra commute with the total excitation number operator $\hat N$, as can be seen by writing a possible representation  $\hat{N}=\hat{a}_1^\dag\hat{a}_1+\hat{a}_2^\dag\hat{a}_2$, $\hat{B}_+=\hat{a}_1^\dag\hat{a}_2+\hat{a}_2^\dag\hat{a}_1$, and $\hat{B}_-=i(\hat{a}_1^\dag\hat{a}_2-\hat{a}_2^\dag\hat{a}_1$). This agrees with the well-known property of MZI that preserves the total number of photons injected into the system. This implies that observables of interest for the thermodynamic analysis, such as the Hamiltonian, commute with all elements of the algebra, and are therefore not affected by the Mach-Zehnder unitary transformation. Clearly, we cannot exclude that conceiving a QHE based on more complex algebras can be of any advantage for other purposes. However, here we have chosen to specialize to a QHE based on the SU(1,1) interference, which introduces great advantages in measuring the observables with precisions that beat the SNL.

\section{Conclusion}\label{cn}
In this work we used mathematical tools from quantum thermodynamics and quantum metrology to investigate the performance of an SU$(1,1)$ quantum heat engine. This allowed us to introduce standard concepts of quantum information theory, such as the quantum Fisher information, which play a fundamental role in quantum thermodynamics (see, for example, the thermodynamic uncertainty relations \cite{PhysRevE.99.062126, PhysRevLett.125.050601,PhysRevLett.126.010602}). Thus, we are applying well-established protocols for phase sensitivity in the context of the performance of quantum thermal machines, an approach that has not yet been sufficiently explored.

Our quantum heat engine exploits the outstanding quantum metrological features of the SU(1,1) interferometer in order to improve the precision in the knowledge of output observables at the cost of a reduced efficiency. Considering that the decrease of the efficiency is inevitable in processes characterized by inner friction (in our case arising due to squeezing), the improvement of the sensitivity offers an advantage in the thermodynamic analysis of QHEs, and can also aid the design and analysis of future experimental implementations.

Finally, we applied our model to a specific superconducting transmission line, showing that it can work as a supersensitive SU$(1,1)$ quantum heat engine.
Clearly, we cannot exclude that the use of other engines in circuit QED, as well as other highly flexible quantum platforms such as Bose-Einstein condensates \cite{RevModPhys.71.463,RevModPhys.78.179}, may provide similar solutions with comparable or better efficiency and/or sensitivity.

\section{Acknowledgments}
The authors thank Ken Funo, Haitao Quan, Peter Hänggi, Gershon Kurizki, Paul Menczel, Dario Poletti and Polina Sharapova for valuable comments and feedback.
A.F. thanks the research center RIKEN for the hospitality.
F.K.W., A.F.,
and D.E.B. acknowledge support from the joint project No. 13N15685 ``German Quantum Computer based on Superconducting Qubits (GeQCoS)'' sponsored by the German Federal Ministry of Education and Research (BMBF) under the \href{https://www.quantentechnologien.de/fileadmin/public/Redaktion/Dokumente/PDF/Publikationen/Federal-Government-Framework-Programme-Quantum-technologies-2018-bf-C1.pdf}{framework programme
``Quantum technologies -- from basic research to the market''}. D.E.B. also acknowledges support from the German Federal Ministry of Education and Research via the \href{https://www.quantentechnologien.de/fileadmin/public/Redaktion/Dokumente/PDF/Publikationen/Federal-Government-Framework-Programme-Quantum-technologies-2018-bf-C1.pdf}{framework programme
``Quantum technologies -- from basic research to the market''} under contract number 13N16210 ``SPINNING''.
F.N. is supported in part by:
Nippon Telegraph and Telephone Corporation (NTT) Research, the Japan Science and Technology Agency (JST) [via the CREST Quantum Frontiers program, the Quantum Leap Flagship Program (Q-LEAP), and the Moonshot R\&D Grant Number JPMJMS2061], and the Office of Naval Research (ONR) Global (via Grant No. N62909-23-1-2074).

\bibliography{ref}

\begin{thebibliography}{96}
\bibitem [{not()}]{note}%
  \BibitemOpen
  \href@noop {} {}\bibinfo {note} {To avoid possible misunderstanding, we stress that the thermodynamic compression corresponds to the analog expansion of the spacetime. Vice versa, the thermodynamic expansion corresponds to the compression of the spacetime.}\BibitemShut {Stop}%
\end{thebibliography}

\end{document}